\documentclass[sigconf,nonacm,natbib=true]{acmart}
\AtBeginDocument{%
  }

\usepackage{csquotes}
\usepackage{pbalance}
\usepackage{mdframed}
\usepackage{enumitem}
\usepackage{tabularx}
\usepackage{xcolor}
\usepackage{array}
\usepackage{makecell}
\usepackage{changepage}
\usepackage{pifont}

\makeatletter
\newcommand{\acmrightssize}{\fontsize{8}{9.5}\selectfont}
\newcommand{\firstpagerights}[1]{%
  \begingroup
    \renewcommand\thefootnote{}%
    \footnotetext{%
      \acmrightssize
      \raggedright
      \setlength{\parskip}{0pt}%
      \setlength{\parindent}{0pt}%
      #1%
    }%
    \addtocounter{footnote}{0}%
  \endgroup
}
\makeatother

\newmdenv[
  backgroundcolor=gray!8,
  linecolor=gray!50,
  roundcorner=4pt,
  skipabove=6pt,
  skipbelow=6pt,
  innerleftmargin=8pt,
  innerrightmargin=8pt,
  innertopmargin=6pt,
  innerbottommargin=6pt
]{prompt}

\newcolumntype{Y}{>{\centering\arraybackslash}X}
\newcolumntype{W}{>{\centering\arraybackslash}X@{\hspace{0pt}}}

\author{Saber Zerhoudi}
\orcid{0000-0003-2259-0462}
\affiliation{%
  \institution{University of Passau}
  \city{Passau}
  \country{Germany}
}
\email{szerhoudi@acm.org}

\author{Adam Roegiest}
\orcid{0000-0003-1265-8881}
\affiliation{%
  \institution{Zuva}
  \city{Toronto}
  \country{Canada}
}
\email{adam@roegiest.com}

\author{Jelena Mitrovi\'{c}}
\orcid{0000-0003-3220-8749}
\affiliation{%
  \institution{University of Passau}
  \city{Passau}
  \country{Germany}
}
\email{jelena.mitrovic@uni-passau.de}

\author{Michael Granitzer}
\orcid{0000-0003-3566-5507}
\affiliation{%
  \institution{University of Passau}
  \city{Passau}
  \country{Germany}
}
\affiliation{%
  \institution{Interdisciplinary Transformation University Austria}
  \city{Linz}
  \country{Austria}
}
\email{michael.granitzer@uni-passau.de}

\begin{document}

\title{As We May Search}
\subtitle{Local-First Information Retrieval}

\begin{abstract}
%Personal documents, legal files, and medical records are valuable search targets, yet current retrieval-augmented generation systems require sending content to remote servers. 
The sensitive information in personal documents, legal files, and medical records is among the most valuable things to search, yet current retrieval-augmented generation systems still require sending content to remote servers. We propose \emph{local-first IR}, a design philosophy where indexes, models, and inference reside on user devices, treating remote services as optional. This paper makes four contributions: (1)~a framework organizing retrieval architectures along three dimensions: privacy and control, capability, and accessibility, (2)~experiments on consumer hardware across five benchmarks, scaling from 1K to 1M documents with dense retrieval, BM25, and hybrid fusion. Dense retrieval keeps over 91\% nDCG@10 up to 100K documents, with approximate HNSW indexes extending this to 1M with only 2\% quality loss; a 7B local language model reaches within 4 points of a cloud baseline on answer quality, (3)~competing perspectives for and against local-first IR, informed by experimental evidence, and (4)~a research agenda identifying open problems. The real tradeoff is scope rather than quality: what matters is what you can search, not how well you can search it.
\end{abstract}

\begin{CCSXML}
<ccs2012>
   <concept>
       <concept_id>10002951.10003317.10003338</concept_id>
       <concept_desc>Information systems~Retrieval models and ranking</concept_desc>
       <concept_significance>500</concept_significance>
       </concept>
   <concept>
       <concept_id>10002951.10003317.10003359</concept_id>
       <concept_desc>Information systems~Evaluation of retrieval results</concept_desc>
       <concept_significance>500</concept_significance>
       </concept>
   <concept>
       <concept_id>10002951.10003260.10003261.10003263</concept_id>
       <concept_desc>Information systems~Web search engines</concept_desc>
       <concept_significance>500</concept_significance>
       </concept>
   <concept>
       <concept_id>10002951.10003317.10003331.10003271</concept_id>
       <concept_desc>Information systems~Personalization</concept_desc>
       <concept_significance>300</concept_significance>
       </concept>
   <concept>
       <concept_id>10002978.10002991.10002995</concept_id>
       <concept_desc>Security and privacy~Privacy-preserving protocols</concept_desc>
       <concept_significance>300</concept_significance>
       </concept>
 </ccs2012>
\end{CCSXML}

\ccsdesc[500]{Information systems~Retrieval models and ranking}
\ccsdesc[500]{Information systems~Evaluation of retrieval results}
\ccsdesc[500]{Information systems~Web search engines}
\ccsdesc[300]{Information systems~Personalization}
\ccsdesc[300]{Security and privacy~Privacy-preserving protocols}

\keywords{Information retrieval, local-first IR, on-device search, privacy-preserving retrieval, retrieval-augmented generation}

\maketitle
\firstpagerights{%
  \textcopyright{} 2026, held by the owner/author(s); published under a Creative Commons Attribution 4.0 International (CC BY 4.0) License.\\
  This is the author's version of the work. The Version of Record was published in the \emph{Proceedings of the 2026 International ACM SIGIR Conference on Innovative Concepts and Theories in Information Retrieval (ICTIR~'26), July 25, 2026, Melbourne, VIC, Australia}, and is available online at: \url{https://doi.org/10.1145/3805713.3820402}.
}

% ══════════════════════════════════════════════════════════════════════
%   INTRODUCTION
% ══════════════════════════════════════════════════════════════════════

\section{Introduction}
\label{sec:intro}

In 1945, Vannevar Bush described a device he called the Memex: ``\emph{a sort of mechanized private file and library}'' in which ``\emph{an individual stores all his books, records, and communications, and which is mechanized so that it may be consulted with exceeding speed and flexibility. It is an enlarged intimate supplement to his memory}''~\citep{bush1945memex}. Bush envisioned a personal device on a desk, not a shared utility.
% The user owns the collection, the index, and the trails through them.

\begin{figure}[t]
\centering
\includegraphics[width=\columnwidth]{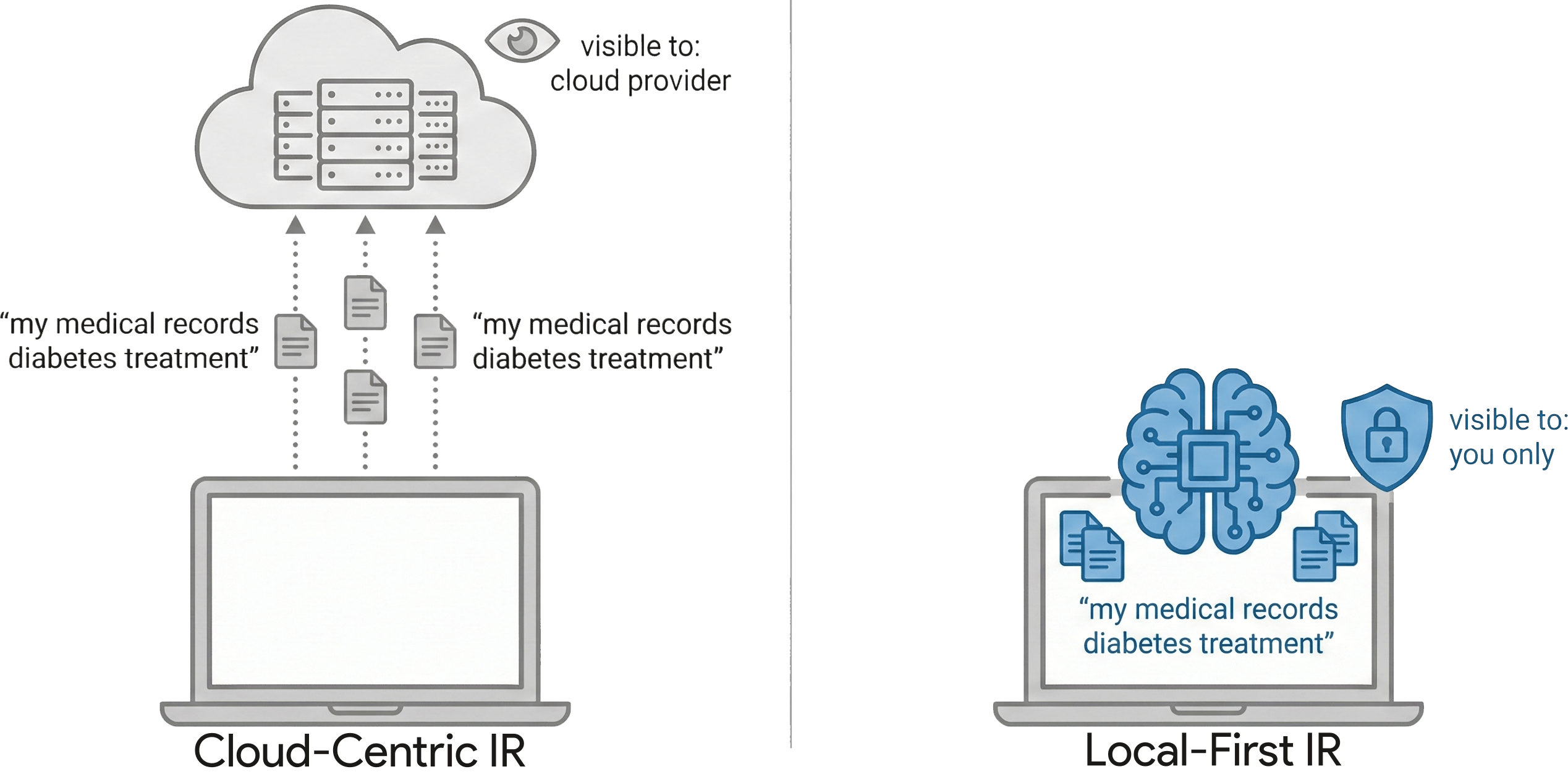}
\caption{Cloud-centric IR (left) sends queries and documents to remote servers visible to operators. Local-first IR (right) processes everything on-device, transmitting zero tokens externally. Users can optionally reach into remote indexes through privacy-preserving hybrids (Section~\ref{sec:hybrid}).}
\label{fig:cloud-vs-local}
\vspace{-2em}
\end{figure}

Eighty years later, the problem Bush identified, the growing impossibility of navigating one's own accumulated knowledge, is more profound than ever. Andrej Karpathy recently framed the same idea as building a ``second brain''~\footnote{\href{https://gist.github.com/karpathy/442a6bf555914893e9891c11519de94f}{Andrej Karpathy, ``LLM Wiki'', GitHub Gist (April 4, 2026).}}, a phrase now common among practitioners. The tools we have built are powerful but architecturally inverted: retrieval-augmented generation (RAG) systems such as ChatGPT~\citep{openai2024chatgpt}, Perplexity~\citep{perplexity2025}, and NotebookLM~\citep{google2025notebooklm} can index personal documents, follow their associative trails, and synthesize answers, but on someone else's server (Figure~\ref{fig:cloud-vs-local}). Every query reveals intent; every uploaded document enters infrastructure the user does not control~\citep{zeng2024ragprivacy,bodea2026sok}. 
% The Memex was private by design; its modern descendants are private only by policy.

The technology to change this has arrived: small, effective language models that run on consumer hardware within mobile power budgets~\citep{apple_foundation_models,chrome_summarizer_gemini_nano}. Embedding models purpose-built for on-device search fit under 200MB of RAM~\citep{google_gemma3n_edge}. Inference frameworks have proliferated: Ollama~\footnote{\url{https://ollama.com}} has served over 52 million monthly downloads, and the Hugging Face GGUF catalog has grown to roughly 170{,}000 models since 2023~\footnote{Hugging Face Hub, models filtered by the GGUF tag, \url{https://huggingface.co/models?library=gguf}, accessed April 15, 2026.}. Millions of users already run local inference daily, yet the IR research community lacks a framework for reasoning about when local retrieval is viable, what its limits are, or how personal knowledge systems interact with the world.

% Desktop search systems demonstrated local indexing two decades ago, but were limited to keyword matching over metadata.
The idea of searching personal files locally is not new. Spotlight, Windows Search, grep, and email search all work offline, but are often limited to keyword matching over metadata (Section~\ref{sec:desktop}).
%Desktop search systems such as Google Desktop, Windows Desktop Search, and Spotlight~\citep{dumais2003stuff,cutrell2006fast} demonstrated local indexing two decades ago,
The personal information management community has advocated for it just as long~\citep{jones2007personal,whittaker2011personal}. What has changed is capability: today's local models support dense semantic retrieval, hybrid fusion, and generative answer synthesis on consumer hardware.

\emph{Local-first IR} addresses this gap: a design philosophy where indexes, models, and inference reside on user devices, treating remote services as optional rather than required. Inspired by local-first software~\citep{kleppmann2019localfirst} and rooted in personal knowledge management~\citep{jones2007personal,bush1945memex}, this approach asks a different question than conventional IR. Instead of ``\emph{how do we build the best search engine?}'' it asks: ``\emph{if a user wants their own searchable knowledge base of the world, as they have encountered it, what is possible and where does it break down?}''

To answer this, we make four contributions:
\begin{enumerate}[label=(\arabic*),labelindent=0pt,leftmargin=2em,labelsep=0.4em]
    \item A \emph{framework} organizing retrieval architectures along three dimensions (privacy, capability, and accessibility) that clarifies what should stay local, the benefits from remote augmentation, and where the cloud remains essential (Section~\ref{sec:design-space}).
    % \item \emph{Experiments at scale} on consumer hardware across five benchmarks, scaling from 1,000 to 1,000,000 documents, comparing dense neural retrieval (sentence-transformers + FAISS), lexical retrieval (BM25), and hybrid fusion, with approximate indexes (HNSW, IVF) evaluated at the largest scales (Section~\ref{sec:experiments}). 
    \item \emph{Experiments at scale} on consumer hardware across five benchmarks, from 1K to 1M documents (Section~\ref{sec:experiments}). We compare dense retrieval, BM25, and hybrid fusion, evaluate approximate indexes (HNSW, IVF) at the largest scales, measure local RAG generation against a cloud baseline, and replicate the scaling curves on an entry-level x86 laptop. A browser-native prototype shows the most constrained deployment (zero-install, cross-platform) is already feasible with current web standards.
    \item An investigation of \emph{how local systems can safely reach into the wider world} through privacy-preserving hybrids, including embedding-only transmission that retains 99.9\% retrieval quality with zero token exposure (Section~\ref{sec:hybrid}).
    \item \emph{A local-first research agenda} that motivates further research to overcome the limitations of current technologies as well as exploring how to best serve users (Sections~\ref{sec:perspectives} and ~\ref{sec:future}).
\end{enumerate}

Our experiments\footnote{Code, data, and prototype: \url{https://github.com/searchsim-org/ictir26-localfirst-ir}} indicate that for personal knowledge bases up to 100,000 documents, local-first retrieval achieves over 91\% of remote-equivalent quality with zero privacy exposure. Collections exceeding that limit trade off retrieval quality and latency. The constraint is scope, what the user can store and index locally, rather than retrieval quality. This paper investigates how far local search can take us before we need the cloud, and how to reach the cloud without surrendering what makes local search valuable.

As Bush wrote: ``\emph{The inheritance from the master becomes, not only his additions to the world's record, but for his disciples the entire scaffolding by which they were erected}''. Local-first IR is about giving users ownership of that scaffolding.

% ══════════════════════════════════════════════════════════════════════
%   RELATED WORK
% ══════════════════════════════════════════════════════════════════════

\section{Related Work}
\label{sec:background}

Building personal knowledge systems with local retrieval requires advances across on-device models, privacy-aware architectures, and accessible inference runtimes. We trace the lineage from early personal search through current on-device inference to identify where the gap we address remains.

\subsection{From Desktop Search to Personal Knowledge Systems}
\label{sec:desktop}

The idea of searching personal files locally has a long history. Bush's Memex~\citep{bush1945memex} proposed a personal, extensible knowledge store where users build associative trails through information. Decades later, desktop search systems realized a subset of this vision: Google Desktop (2004), Windows Desktop Search~\citep{cutrell2006fast}, and Apple Spotlight provided keyword indexing over local files, while Stuff I've Seen~\citep{dumais2003stuff} demonstrated re-finding based on interaction history. The personal information management (PIM) community has studied how people organize, retrieve, and re-find their own information~\citep{jones2007personal,whittaker2011personal,elsweiler2007towards,zerhoudi2024generative}. These systems demonstrate that users value local search, but were limited to lexical matching over metadata and file contents. Modern RAG systems close the capability gap with semantic retrieval and generative synthesis, but reintroduce centralized control. Others reduce this dependence: open web indexes query a shared index without a central engine~\citep{hendriksen2026open}, and agentic systems may use sandboxed retrieval environments in place of live web search~\citep{sun2025zerosearch}.

%Local-first IR returns to the desktop search premise with stronger tools: dense retrieval, hybrid fusion, and on-device language models.

\subsection{On-Device Models and Inference}

Industry has moved small language models (SLMs) from research to production. Chrome ships Gemini Nano for local summarization~\citep{chrome_summarizer_gemini_nano}, Apple Intelligence runs foundation models on-device~\citep{apple_foundation_models}, and Google's Gemma~3n targets edge inference~\citep{google_gemma3n_edge}, all building on 1--4B parameter models such as Phi-4~\citep{abdin2024phi3}, Llama~3.2~\citep{meta2024llama32}, and SmolLM2~\citep{benallal2025smollm2}. On the retrieval side, embedding models like MiniLM~\citep{microsoft_minilm} enable semantic search under 200MB of RAM. Browser runtimes including WebLLM~\citep{ruan2024webllm} and MeMemo~\citep{wang2024mememo} show that neural retrieval and generation pipelines can run inside a browser tab, though benchmarks reveal persistent overhead versus native execution~\citep{wang2025anatomizing}. Knowledge distillation~\citep{hinton2015distilling}
%, which compresses a large model into a smaller one while preserving capability, 
is a useful tool for local-first IR because it moves frontier capabilities to the user's machine. Local-first IR is not about training individual models, but about where and how the full retrieval pipeline executes.

% Local-first IR differs from knowledge distillation~\citep{hinton2015distilling} in a fundamental way. Distillation compresses a large model into a smaller one that runs locally, preserving as much capability as possible. Local-first IR is an architectural philosophy about where the entire retrieval pipeline executes, not about how individual models are trained. A local-first system may use distilled models, but it may equally use models trained from scratch for on-device deployment. Our contribution is the framework and evaluation, not a new training method.

\subsection{Privacy-Preserving Retrieval}

RAG pipelines expose user data at multiple stages: queries reveal intent, retrieved passages expose corpus content, and generated responses may leak information~\citep{zeng2024ragprivacy}. Existing mitigations retain remote dependencies. Federated systems~\citep{fedrag,hyfedrag} distribute computation but require coordination servers. Confidential computing~\citep{apple2024pcc,addison2024cfedrag} isolates processing via hardware enclaves (Trusted Execution Environments). Differential privacy~\citep{dprag} provides formal leakage bounds but with steep utility costs. Homomorphic encryption enables computation on encrypted data without decryption~\citep{gentry2009fully}, though the computational overhead currently makes it impractical for interactive retrieval. These approaches share a common thread: they mitigate exposure during remote processing rather than eliminating remote processing entirely. Local-first IR takes a complementary position: avoid transmission in the first place, and when external access is needed, transmit only what the index requires, not explicitly what the user queried for (Section~\ref{sec:hybrid}).

\subsection{Local-First Software Principles}
Our work draws on the local-first software movement, where network connectivity is an enhancement rather than a dependency, prioritizing data ownership, offline utility, and longevity~\citep{kleppmann2019localfirst}. The insight reappears wherever connectivity is constrained~\citep{clarke2017mars,andersen2009fawn}. We formalize this as ``transfer-once, interact-fast'': download models and indexes once, then operate locally. The ``slow search'' paradigm suggests users accept this bargain when the payoff is clear~\citep{teevan2013slow}.

\subsection{Gap and Contribution}
The pieces exist (on-device models, local vector search, privacy-preserving protocols), but two gaps remain. The \emph{model gap}: on-device inference today serves single tasks (summarization, rewriting), but no system connects these into a complete retrieval pipeline evaluated at meaningful scale. The \emph{scaling gap}: prior desktop search and recent browser-based demonstrations operate at personal scale (hundreds to low thousands of documents) but have not characterized how quality, latency, and cost evolve as the knowledge base grows toward the hundreds of thousands or millions of documents that real personal and organizational collections contain. We address both by framing local-first IR as a design philosophy, backed by experiments scaling from 1,000 to 1,000,000 documents across three retrieval methods, three FAISS index types, and five local generation models on consumer hardware. The contribution is the framework and study, not a new algorithm.

% ══════════════════════════════════════════════════════════════════════
%   DESIGN SPACE
% ══════════════════════════════════════════════════════════════════════

\section{The Local-First IR Design Space}
\label{sec:design-space}

Building a personal knowledge base requires choices about where data lives, what capabilities are available, and how it is deployed. Keeping personal files in a shared office gives access to copiers, filing systems, and colleagues, but everyone can potentially see your documents\footnote{This happened in the Samsung data leak in 2023~\cite{park2023samsung}.}. Keeping them at home means a smaller workspace but more privacy. Each architectural choice in this section answers: \emph{how much of the office's infrastructure are you willing to use, and what is revealed by using it?}

\subsection{Three Dimensions}

\paragraph{Privacy and Control.}
\emph{How much control do users retain over their data?} Fully local systems keep indexes, models, queries, and interaction logs on-device. Cloud-centric systems centralize everything. Between these poles: confidential computing uses hardware enclaves~\citep{apple2024pcc}; federated systems share derived signals while retaining raw text~\citep{fedrag,hyfedrag}; homomorphic encryption computes on ciphertext without decryption~\citep{gentry2009fully}; and hybrid architectures perform sensitive operations locally before selectively querying remote services.

\paragraph{Capability.}
\emph{What can the system find and reason about?} Cloud infrastructure supports trillion-parameter models and web-scale indexes. Local systems operate within consumer hardware limits: 1--4B parameter models, corpora bounded by device storage~\citep{abdin2024phi3,meta2024llama32}. Local systems cannot index the open web or match cloud-scale reasoning, and users can only find what they already possess. But for the knowledge a user has accumulated over a career, a case, or a diagnosis, the gap is smaller than one might assume.

\paragraph{Accessibility.}
\emph{How easily can users reach the system?} Cloud APIs work from any browser but require payment and trust. Native applications use full hardware but require installation. Browser-native systems run cross-platform without installation and work offline after first load, constrained only by browser support~\citep{w3c2023webgpu,mdn2024opfs}. These categories are not fixed: a system might default to local operation but escalate to cloud when the user's need exceeds local capability (with user permission).

\subsection{Architectural Positions}

\begin{table*}[t]
\captionsetup{skip=4pt}
\centering
\caption{The Local-First IR Design Space. Architectures vary in where data resides, what capabilities they offer, and how accessible they are to deploy. The rightmost columns indicate which components execute locally vs.\ remotely for each position.}
\label{tab:design-space}
\small
\setlength{\tabcolsep}{4pt}
\renewcommand{\arraystretch}{1.08}

\begin{tabularx}{\textwidth}{@{}
p{0.18\textwidth}
>{\centering\arraybackslash\hsize=1.4\hsize}X
>{\centering\arraybackslash\hsize=1\hsize}X
>{\centering\arraybackslash\hsize=0.8\hsize}X
>{\centering\arraybackslash\hsize=0.8\hsize}X
|
*{3}{>{\centering\arraybackslash}X}
@{}}

\toprule
& \multicolumn{4}{c|}{\textit{Design Dimensions}} & \multicolumn{3}{c}{\textit{Pipeline Execution}} \\
\cmidrule(lr){2-5} \cmidrule(l){6-8}
\textbf{Architecture} &
\textbf{Privacy \& Control} &
\textbf{Capability} &
\textbf{Accessibility} &
\textbf{Offline} &
\textbf{Retrieval} &
\textbf{Ranking} &
\textbf{Generation} \\
\midrule
Cloud RAG API &
Low & High & High & No & Remote & Remote & Remote \\
Confidential Computing &
Medium & High & Medium & No & Remote (TEE) & Remote (TEE) & Remote (TEE) \\
Federated RAG &
Medium & Medium--High & Low & Partial & Distributed & Distributed & Varies \\
Hybrid Local &
Medium--High & Medium & High & Partial & Remote & Local & Local \\
\textbf{Fully Local} &
\textbf{High} & \textbf{Low--Medium} & \textbf{High} & \textbf{Yes} &
\textbf{Local} & \textbf{Local} & \textbf{Local} \\
\bottomrule
\end{tabularx}
\end{table*}

Table~\ref{tab:design-space} maps five representative architectures onto this space. \textbf{Cloud RAG APIs} maximize capability at the cost of privacy. \textbf{Confidential Computing} balances scalability with hardware-enforced isolation~\citep{apple2024pcc}. \textbf{Federated RAG} distributes computation to keep raw corpora local~\citep{fedrag}. \textbf{Hybrid Local} performs sensitive operations on-device and reaches remote indexes only through sanitized representations~\citep{pathak2025RAG,ye2026efficient}, though transmitting embeddings carries risks~\citep{morris2023embedding}. \textbf{Fully Local} executes the entire pipeline on-device, maximizing privacy and offline capability at the cost of scope. Availability ratings assume connectivity; offline, only the fully local row survives.

These five positions form a continuum that systems navigate dynamically rather than a set of competing alternatives. A personal knowledge system should default to fully local operation and escalate to hybrid or remote only when the user's information need exceeds local boundaries, with escalation requiring explicit consent.

\subsection{Design Principles}

One might argue that the decision is simpler than a three-dimensional framework suggests: private data must stay local, non-private data goes to the cloud, and there are no additional choices~\citep{bodea2026sok}. In many deployment scenarios, this heuristic suffices, and we view it as a starting point. But the boundary between private and non-private is rarely clear in practice. A researcher's literature collection may be public individually yet reveal a confidential research direction in aggregate. A lawyer's query to a public legal database might reveal data around the private case informing the search. The framework adds value precisely where the heuristic breaks down: when users need to reach external resources without revealing why, when operational constraints (e.g., compliance, bandwidth, device capability) shape the choice, and when the system must transition between modes with the user's informed consent.

The design space suggests four principles. \emph{Match architecture to task}: personal search suits local deployment; web-scale needs suit the cloud. \emph{Clarify tradeoffs}: ``on-device'' search that transmits embeddings occupies a different privacy position than fully local. \emph{Enable dynamic transitions}: systems should shift between local and remote, but the shift requires consent. \emph{Amortize transfer costs}: download models and indexes once, then operate locally thereafter~\citep{teevan2013slow}.

% ══════════════════════════════════════════════════════════════════════
%   EXPERIMENTS
% ══════════════════════════════════════════════════════════════════════

\section{How Far Can Local Search Take Us?}
\label{sec:experiments}

% If a user builds a local knowledge base, starting with personal documents and growing to organizational scale, how does search quality, cost, and latency evolve? When does the system remain viable, and where does it break down?

In this section, we explore how a local knowledge base might scale when it grows from modest local collections to organizational ones. We explore how search quality, cost, and latency change as well as when and where the system could break down.

\subsection{Experimental Setup}

\paragraph{Implementation.}
We implement three retrieval methods and two LLM-generation paths as native processes on consumer hardware.
\begin{itemize}[label=\textendash,labelindent=0pt,leftmargin=2em,labelsep=0.4em]
\item \textbf{Dense:} Semantic retrieval using \texttt{\small all-MiniLM-L6-v2} (22M parameters) via sentence-transformers~\citep{reimers2019sentence}, with FAISS~\citep{johnson2019billion} inner-product indexing over 384-dimensional embeddings. We evaluate three index types: exact flat search (\texttt{\small IndexFlatIP}), graph-based approximate search (\texttt{\small IndexHNSWFlat}~\citep{malkov2018hnsw}, $M{=}32$, $\mathit{efConstruction}{=}200$, $\mathit{efSearch}{=}64$), and inverted-file approximate search (\texttt{\small IndexIVFFlat}, $\mathit{nlist}{=}100$, $\mathit{nprobe}{=}10$).
\item \textbf{BM25:} Lexical retrieval via Okapi BM25~\citep{robertson1994bm25} ($k_1{=}1.5$, $b{=}0.75$), pure-Python \texttt{\small rank\_bm25}, no neural model dependencies.
\item \textbf{Hybrid:} Reciprocal Rank Fusion~\citep{cormack2009rrf} (RRF, $k{=}60$) combining dense and BM25 at a 0.7/0.3 weight ratio\footnote{HNSW and IVF settings are FAISS defaults; $k{=}60$ follows the original RRF paper. The 0.7/0.3 ratio favors dense, the stronger method here. None were tuned per corpus.}.
\item \textbf{Local generation:} Dense retrieval followed by answer generation via Ollama, evaluated with five small instruction-tuned, non-reasoning models: \texttt{\small qwen2.5:1.5b}, \texttt{\small gemma4:e4b}, \texttt{\small qwen2.5:3b}, \texttt{\small qwen2.5:7b}, and \texttt{\small gpt-4o-mini} as a remote non-reasoning cloud baseline. We chose non-reasoning variants on both sides so that generation latency is directly comparable; reasoning models would shift the latency picture without changing the answer-quality story we are after.
\end{itemize}

All retrieval and local-generation methods execute entirely on the user's device with zero external network requests. The remote-RAG condition exists only as a reference for comparison; no tokens, queries, or embeddings leave the machine in any local condition.

\paragraph{Datasets.}
We evaluate on three experimental conditions:

%\noindent 
\emph{Scaling condition.} MS~MARCO passage subsets~\citep{nguyen2016msmarco} at eight corpus sizes: 1K, 5K, 10K, 25K, 50K, 100K, 500K, and 1M documents, each with 200 queries and complete relevance judgments\footnote{Human-labeled annotations marking, for each query, which passages contain a correct answer.}. Each subset was independently constructed by including all gold passages for evaluation queries, then filling the remainder with randomly sampled passages until the target size was reached. %Subsets were built independently; smaller sets are not strict subsets of the larger ones.
% The 1M collection was constructed by merging the 500K subset with 500K new passages sampled independently from the MS~MARCO corpus, ensuring that all gold passages for the 200 evaluation queries are present. This condition explores how well local search scales.

\emph{Domain condition.} Four domain-specific benchmarks at standard BEIR sizes~\citep{thakur2021beir}: SciFact (5,183 scientific abstracts, 300 queries)~\citep{wadden2020scifact}, FiQA (10,000 financial passages, 648 queries)~\citep{maia2018fiqa}, Natural Questions (10,000 Wikipedia passages, 300 queries)~\citep{kwiatkowski2019nq}, and TREC-COVID (50,000 medical abstracts, 50 queries)~\citep{roberts2020treccovid}. This condition tests how well local search generalizes across domains.

\emph{Generation condition.} Natural Questions passages (10K) augmented with short-form gold answers from \texttt{\small nq\_open}~\citep{kwiatkowski2019nq}. Of the 3,452 BEIR queries, 2,255 (65.3\%) match a question in \texttt{\small nq\_open} and carry a gold short answer. We take the first 300 queries in corpus order without filtering for gold-answer availability. Of those 300, 147 happen to have a gold answer (consistent with the 65.3\% base rate). All 300 queries run through the full pipeline (retrieval + generation) and contribute latency measurements, but we compute EM, F1, and \emph{contains} only on the 147 that have a gold string to compare against. This condition quantifies how closely several local small language models approximate a cloud baseline on RAG answer quality.

\paragraph{Metrics.}
For retrieval: nDCG@10 (a standard measure of ranking quality), MRR@10, and Recall@10; cold-start time (model loading + index construction), warm-start latency (per-query after initialization), and P50/P95 latency distributions. Peak resident set size (RSS) is tracked via \texttt{\small psutil}. For generation: exact match (EM), token-level F1, and the \emph{contains} rate (if a gold answer appears in the generated text, after case-folding and stop-word removal). Significance tests use two-sided paired t-tests with Cohen's $d$ effect sizes. We report at rank 10 throughout, per TREC convention.

\paragraph{Hardware.}
Our main target is a commodity Apple~M1 laptop with 16\,GB of RAM running macOS. To avoid over-fitting on Apple Silicon, we replicate the same experiments on a second machine: an entry-level x86 laptop running Ubuntu 20.04 with an 8-core CPU, 16\,GB of RAM, no dedicated GPU, and Python 3.10. Running the same pipeline on two platforms separates algorithmic scaling, which should hold anywhere, from absolute timing, which will not.

\subsection{Scaling: From Personal Files to Organizational Archives}

% Table~\ref{tab:scaling} shows how retrieval quality and cold-start cost change as the corpus grows from 1,000 to 1,000,000 documents.

\begin{table}[h]
\captionsetup{skip=0pt}
\centering
\caption{Retrieval quality and cold-start cost from 1K to 1M documents on an Apple~M1 and 200 MS~MARCO queries.}
% \caption{Retrieval quality and cold-start cost as the local knowledge base grows from 1K to 1M documents on the Apple~M1 laptop. Dense retrieval (flat FAISS) loses 8\% nDCG@10 from 1K to 100K, and another 17\% from 100K to 1M. BM25 indexes in seconds at every scale but trails dense in ranking quality.}
\label{tab:scaling}
\small
\setlength{\tabcolsep}{3pt}
\renewcommand{\arraystretch}{1.05}

\begin{tabular*}{\linewidth}{@{\extracolsep{\fill}}r ccc cc@{}}
\toprule
& \multicolumn{3}{c}{\textbf{nDCG@10}} & \multicolumn{2}{c}{\textbf{Cold Start}} \\
\cmidrule(lr){2-4} \cmidrule(lr){5-6}
\textbf{Docs} & \textbf{Dense} & \textbf{BM25} & \textbf{Hybrid} & \textbf{Dense} & \textbf{BM25} \\
\midrule
1K    & 0.990 & 0.835 & 0.958 & 2.4\,s   & 17\,ms \\
5K    & 0.982 & 0.749 & 0.910 & 10.7\,s  & 85\,ms \\
10K   & 0.969 & 0.711 & 0.903 & 22.6\,s  & 163\,ms \\
25K   & 0.955 & 0.653 & 0.887 & 48.8\,s  & 446\,ms \\
50K   & 0.922 & 0.616 & 0.870 & 1.6\,min & 1.7\,s \\
100K  & 0.910 & 0.592 & 0.849 & 3.2\,min & 2.1\,s \\
500K  & 0.789 & 0.448 & 0.734 & 31.6\,min & 10.3\,s \\
1M    & 0.736 & 0.409 & 0.680 & 47.5\,min & 19.0\,s \\
\bottomrule
\end{tabular*}
\par\vspace{3pt}
% \footnotesize Dense P50 query latency: 9\,ms (1K) to 41\,ms (1M) with a flat FAISS index. BM25 P50 grows from 1\,ms at 1K to 852\,ms at 1M and dominates hybrid latency above 100K. Peak resident memory: 624\,MB (1K) to 4.9\,GB (1M) for dense. All numbers are measured on 200 MS~MARCO queries per scale.
\footnotesize Peak resident memory: 624\,MB (1K) to 4.9\,GB (1M) for dense.
\end{table}

\paragraph{Quality degrades gracefully.} From Table~\ref{tab:scaling}, dense retrieval (flat index) loses 8\% nDCG@10 from 1K to 100K (0.990 to 0.910). At 500K, well beyond typical personal collections~\citep{whittaker2011personal}, nDCG@10 remains 0.789. At 1M it drops to 0.736, still above BM25 (0.409) and hybrid (0.680) at the same scale. Recall@10 stays above 0.9 at every scale below 100K: the system still finds the relevant documents, it just ranks them less precisely as more plausible distractors appear. The dense-over-BM25 gap is statistically significant: paired t-tests at every scale give $p < 10^{-5}$ with Cohen's $d$ between $+0.45$ (5K) and $+0.73$ (500K), significant and moderate-to-large in effect size.

\paragraph{Embedding is the expensive step, not indexing.} Dense cold start is 3.2\,minutes at 100K, 31.6\,minutes at 500K, and 47.5\,minutes at 1M. Over 99\% of this is the embedding pass. The FAISS flat index build itself is under 1\,second at 1M once embeddings exist (measured against the disk cache described in Section~\ref{sec:ann}). BM25 builds the same three corpora in 2.1\,seconds, 10.3\,seconds, and 19.0\,seconds, so the ratio against dense cold start is roughly 90$\times$ at 100K and 150$\times$ at 1M. The practical deployment recipe follows: serve BM25 immediately while dense embeddings compute in the background, and disk-cache the embeddings so later rebuilds with different index types take seconds. 
%The actual bottleneck is the embedding model itself. Faster or parallel small embedders would move this line further than any ANN parameter choice.

% Dense indexing takes 3.2\,minutes at 100K documents, 31.6\,minutes at 500K, and 47.5\,minutes at 1M. BM25 indexes the same three scales in 2.1\,seconds, 10.3\,seconds, and 19.0\,seconds. The ratio is roughly 90$\times$ at 100K and 150$\times$ at 1M. A practical deployment strategy follows directly: start with BM25 for instant search and compute dense embeddings in the background, then swap in the dense index when it is ready. We also disk-cache the computed embeddings so that a subsequent rebuild with a different index type (Section~\ref{sec:ann}) takes seconds instead of tens of minutes.

\paragraph{BM25 latency is a pure-Python bottleneck.} The BM25 numbers above are for \texttt{\small rank\_bm25}, a reference Python implementation. Per-query latency grows linearly with corpus size: 1\,ms at 1K, 452\,ms at 500K, and 852\,ms at 1M. At scale this influences hybrid retrieval performance, whose latency is dominated by BM25. A native engine such as PISA~\citep{mallia2019pisa}, pyserini~\citep{lin2021pyserini}, or Tantivy~\footnote{\url{https://github.com/quickwit-oss/tantivy}} would be one to two orders of magnitude faster and change the latency picture, but not the quality picture. We report the Python numbers to reflect what a user pip-installing an off-the-shelf pipeline would experience.

\paragraph{Dense query latency stays interactive through 1M} Dense P50 goes from 9\,ms at 1K to 41\,ms at 1M on the flat index, P95 tops out at 76\,ms. Every scale stays within the 100\,ms budget that keeps search feeling instant. BM25 sits at sub-millisecond up to around 100K, then climbs to 452\,ms at 500K and 852\,ms at 1M because the Python scorer is $O(N)$ per query. The cost that hurts users is never the query; it is the embedding pass that precedes it.
%Faster or more parallel small embedding models would move this line more than any ANN parameter choice.

\begin{figure}[t]
\centering
\includegraphics[width=\columnwidth]{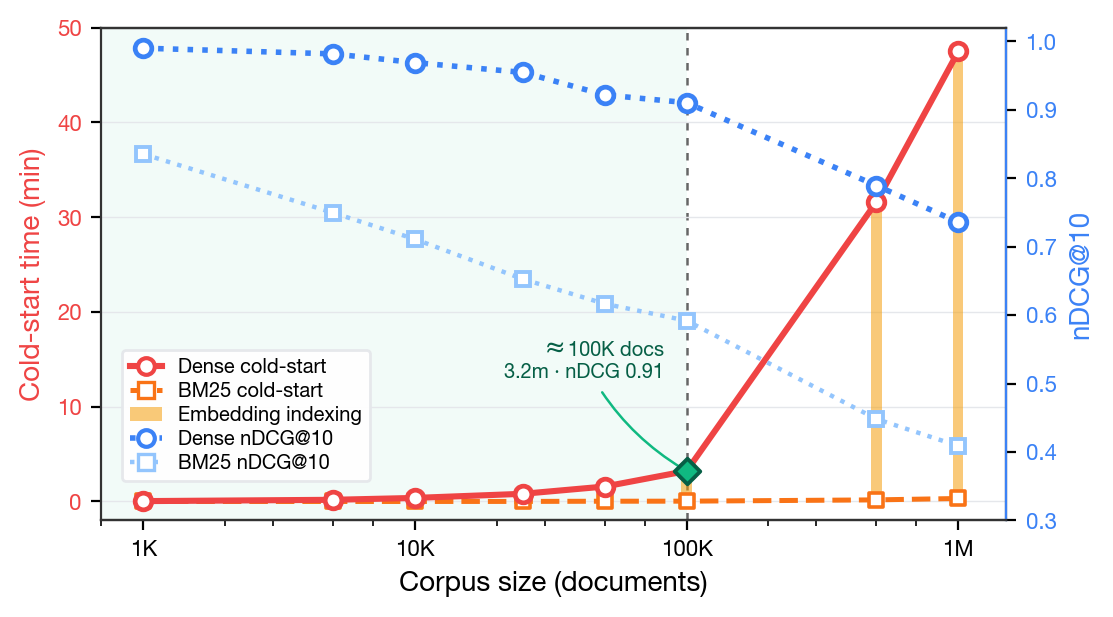}
\caption{Retrieval quality (nDCG@10, right axis) and cold-start time (left axis) from 1K to 1M documents. Below 100K, dense retrieval holds above 0.91 nDCG@10 with cold start under a few minutes. Beyond 100K, cold start grows sharply. BM25 indexes in seconds at every scale but trails dense.}
% \caption{Retrieval quality (nDCG@10, right axis) and cold-start time (left axis) for dense and BM25 retrieval from 1K to 1M documents.}
\label{fig:scaling-curve}
\end{figure}

Figure~\ref{fig:scaling-curve} summarizes these results across three zones. Below 10K documents, which covers personal files and project notes, cold start is under 30 seconds and quality is above 0.97. Between 10K and 100K, which covers a researcher's career output or a small departmental archive~\footnote{Dinneen et al.\ found that personal file collections of knowledge workers typically contain 30{,}000--190{,}000 files~\cite{dinneen2019scale}, placing a researcher's career output in this range.}, cold start reaches 3.2 minutes with quality above 0.91, an acceptable wait that amortizes across every subsequent query and supports incremental additions as new documents arrive. Beyond 100K, cold start jumps from 3 minutes at 100K to 32 minutes at 500K and 48 minutes at 1M with the exact flat index. Section~\ref{sec:ann} shows that approximate indexes recover interactive cold start at the 1M scale, and Section~\ref{sec:hybrid} shows how to combine local indexing with selective remote scope when the corpus outgrows the device.

\subsection{Scaling Further with Approximate Indexes}
\label{sec:ann}

Flat inner-product search is exact but expensive to build. In contrast, HNSW and IVF trade small amounts of recall for large build-time and query-time speedups. Table~\ref{tab:ann} reports the quality and cost of all three index types at 500K and 1M documents, with cold start split into its two components: embedding the corpus, and building the FAISS structure on top of the embeddings.
%HNSW builds a layered proximity graph. IVF clusters the corpus into coarse cells that the query scans selectively. Table~\ref{tab:ann} reports the changes when using an approximate index. Note that these experiments reused disk-cached embeddings from the flat run, so the reported cold start covers only the FAISS index build, not the encoding step.

\begin{table}[t]
\captionsetup{skip=0pt}
\centering
\caption{Flat versus HNSW versus IVF at 500K and 1M documents, with cold start split into embedding and index-build components. HNSW reaches sub-15\,ms P50 query latency at 1M with a 2\% nDCG@10 loss; its index build from cached embeddings takes 3.2 minutes at 1M, versus seconds for IVF.}
\label{tab:ann}
\small
\setlength{\tabcolsep}{3pt}
\renewcommand{\arraystretch}{1.05}
\begin{tabular*}{\linewidth}{@{\extracolsep{\fill}}r l c c c c c@{}}
\toprule
\textbf{Docs} & \textbf{Index} & \textbf{nDCG@10} & \textbf{Recall@10} & \textbf{P50} & \textbf{Embed} & \textbf{Build} \\
\midrule
500K & Flat & 0.789 & 0.892 & 27\,ms & 31.6\,min & 0.4\,s \\
500K & HNSW & 0.777 & 0.877 & 15\,ms & 31.6\,min & $\sim$1\,min \\
500K & IVF  & 0.763 & 0.858 & 15\,ms & 31.6\,min & 0.7\,s \\
\addlinespace
1M   & Flat & 0.736 & 0.868 & 41\,ms & 47.5\,min & 0.8\,s \\
1M   & HNSW & 0.719 & 0.833 & 11\,ms & 47.5\,min & 3.2\,min \\
1M   & IVF  & 0.677 & 0.790 & 15\,ms & 47.5\,min & 1.5\,s \\
\bottomrule
\end{tabular*}
\par\vspace{3pt}
% \footnotesize P50 over 200 queries. Embed is the time to compute all document embeddings with \texttt{all-MiniLM-L6-v2}, identical across index types at a given scale and cached to disk after the first run. Build is the FAISS index construction time on cached embeddings; switching index types later pays only this column. HNSW: $M{=}32$, $\mathit{efConstruction}{=}200$, $\mathit{efSearch}{=}64$. IVF: $\mathit{nlist}{=}100$, $\mathit{nprobe}{=}10$.
\footnotesize HNSW: $M{=}32$, $\mathit{efConstruction}{=}200$, $\mathit{efSearch}{=}64$. IVF: $\mathit{nlist}{=}100$, $\mathit{nprobe}{=}10$.
\end{table}

Two findings matter for a local-first system. HNSW at 1M hits 11\,ms P50, almost four times faster than the 41\,ms of the flat index at the same scale, for a 2\% nDCG@10 cost. And the embedding cache reduces the wait from 47 minutes to 3 minutes whenever the user switches index types or recovers from a crash. Amortizing the embedding pass across index variants is a simple, large win.

The practical recipe therefore follows three tiers. Serve BM25 first, since it is instant at every scale and gives the user something to search against while the embedding pass runs. As soon as the embeddings are cached, build an IVF index (under 2 seconds at 1M) to switch the interface to dense retrieval right away. In parallel, build HNSW for long-term serving; when it is ready, swap it in for the IVF index to recover 4\% nDCG@10 and drop P50 latency from 15\,ms to 11\,ms. At steady state the user runs on HNSW, and BM25/IVF merely exist to shorten the time to first dense result.
%IVF is a third option with different tradeoffs: its index build is near-instant (under 2\,seconds at 1M from a warm cache), so it is a good fit for fast iteration and for cases where the corpus changes often, at the cost of about 6\% additional nDCG@10 loss compared to the flat baseline.

\subsection{Cross-Platform Replication}
\label{sec:xplatform}
As users have different hardware, we check result consistency by replicating dense, BM25, and hybrid at 10K, 100K, and 500K documents on the entry-level x86 laptop described in Section~\ref{sec:experiments} (8-core CPU, 16\,GB RAM, no GPU, Ubuntu 20.04), under the same 200-query evaluation protocol (Table~\ref{tab:xplatform}).

\begin{table}[H]
\captionsetup{skip=0pt}
\centering
% \caption{Cross-platform replication on the entry-level x86 laptop (8-core CPU, 16\,GB RAM, no GPU, Ubuntu 20.04). Retrieval quality matches the Apple~M1 numbers to three decimal places at every scale, which is the expected behavior of deterministic algorithms on identical data. Warm-query latency is close on both platforms. Cold start is 4 to 5 times slower on x86 because the CPU lacks the ML-specific kernel acceleration present in Apple Silicon.}
\caption{Cross-platform replication on the entry-level x86 laptop (8-core CPU, 16\,GB RAM, no GPU, Ubuntu 20.04). Quality matches Apple~M1 (deterministic algorithms, identical data). Warm latency is close on both. Cold start is 4--5$\times$ slower on x86 due to lack of ML-specific kernel acceleration.}
\label{tab:xplatform}
\small
\setlength{\tabcolsep}{3pt}
\renewcommand{\arraystretch}{1.05}

\begin{tabular*}{\linewidth}{@{\extracolsep{\fill}}r l c c c c@{}}
\toprule
& & \multicolumn{2}{c}{\textbf{Apple M1}} & \multicolumn{2}{c}{\textbf{x86 Laptop}} \\
\cmidrule(lr){3-4}\cmidrule(lr){5-6}
\textbf{Docs} & \textbf{Method} & \textbf{nDCG@10} & \textbf{Cold} & \textbf{nDCG@10} & \textbf{Cold} \\
\midrule
10K  & Dense  & 0.969 & 22.6\,s  & 0.969 & 107.6\,s \\
10K  & BM25   & 0.711 & 0.2\,s   & 0.711 & 0.3\,s \\
10K  & Hybrid & 0.903 & 6.1\,s   & 0.903 & 2.6\,s \\
\addlinespace
100K & Dense  & 0.910 & 3.2\,min & 0.910 & 18.1\,min \\
100K & BM25   & 0.592 & 2.1\,s   & 0.592 & 2.6\,s \\
100K & Hybrid & 0.849 & 4.2\,s   & 0.849 & 5.4\,s \\
\addlinespace
500K & Dense  & 0.789 & 31.6\,min & 0.789 & 94.6\,min \\
500K & BM25   & 0.448 & 10.3\,s  & 0.448 & 14.0\,s \\
500K & Hybrid & 0.734 & 13.9\,s  & 0.734 & 17.1\,s \\
\bottomrule
\end{tabular*}
\end{table}

The nDCG@10 values match at every scale for every method, as expected from deterministic algorithms on identical data. What changes across platforms is latency. Dense cold start on the x86 laptop is 4.8$\times$ slower at 10K and 3.0$\times$ slower at 500K than on the M1. Warm P50 stays within the same tens-of-milliseconds band on both. The takeaway: the scaling shape is the same on both platforms and the thresholds from Section~\ref{sec:experiments} are platform-neutral. Users with similar x86 hardware should expect the same curves, shifted by a small factor on cold start and unchanged on warm latency.

\subsection{Cross-Domain Evaluation}

A personal knowledge base is rarely homogeneous. A clinician collects medical literature; a financial analyst accumulates market reports; a researcher spans multiple fields. Table~\ref{tab:domain} tests whether local search generalizes across domains.

\begin{table}[h]
\captionsetup{skip=0pt}
\centering
\caption{Cross-domain retrieval effectiveness (nDCG@10) across four BEIR datasets.
%BM25 achieves 48--85\% of dense depending on domain specificity.
}
\label{tab:domain}
\small
\setlength{\tabcolsep}{3pt}
\renewcommand{\arraystretch}{1.05}

\begin{tabularx}{\linewidth}{@{}p{0.22\linewidth} r *{4}{>{\centering\arraybackslash}X}@{}}
\toprule
\textbf{Dataset} & \textbf{Docs} & \textbf{Dense} & \textbf{BM25} & \textbf{Hybrid} & \textbf{BM25/Dense} \\
\midrule
SciFact      & 5K  & 0.640 & 0.544 & \textbf{0.651} & 85\% \\
FiQA         & 10K & \textbf{0.545} & 0.263 & 0.512 & 48\% \\
Nat.\ Questions & 10K & \textbf{0.888} & 0.583 & 0.833 & 66\% \\
TREC-COVID   & 50K & 0.784 & 0.555 & \textbf{0.803} & 71\% \\
\bottomrule
\end{tabularx}
\par\vspace{3pt}
\footnotesize Bold indicates best condition. Dense-vs-BM25 differences are significant ($p < 0.001$).
\end{table}

\paragraph{Dense retrieval is essential for domain-specific corpora.} On FiQA (financial), BM25 achieves only 48\% of dense nDCG@10 because specialized vocabulary and semantic relationships require neural representations. On general-purpose corpora (Natural Questions), the gap narrows but dense still leads significantly.

\paragraph{Hybrid improves on domain-specific collections.} On SciFact and TREC-COVID, hybrid RRF \emph{outperforms} pure dense retrieval, because lexical matching captures domain-specific terminology (drug names, clinical codes) that the general-purpose embedding model misses. On NQ and FiQA, the BM25 component dilutes strong dense signals, and pure dense wins.

\paragraph{The practical recommendation.} For a personal knowledge base spanning multiple domains, dense retrieval with optional BM25 fusion is reasonable. BM25 alone suffices for general-purpose corpora and serves as a fallback while the dense index builds.

\subsection{Latency and the User Experience}

\begin{table}[h]
\captionsetup{skip=0pt}
\centering
\caption{Cross-domain latency. Dense P50 stays under 30\,ms on every benchmark; cold start depends on corpus size.}
\label{tab:latency}
\small
\setlength{\tabcolsep}{3pt}
\renewcommand{\arraystretch}{1.05}

\begin{tabular*}{\linewidth}{@{\extracolsep{\fill}}l l r r r@{}}
\toprule
\textbf{Dataset} & \textbf{Method} & \textbf{Cold Start} & \textbf{P50} & \textbf{P95} \\
\midrule
SciFact    & Dense & 31\,s    & 11\,ms  & 60\,ms \\
           & BM25  & 0.5\,s   & 12\,ms  & 26\,ms \\
\addlinespace
FiQA       & Dense & 55\,s    & 13\,ms  & 44\,ms \\
           & BM25  & 0.5\,s   & 25\,ms  & 54\,ms \\
\addlinespace
NQ         & Dense & 41\,s    & 13\,ms  & 42\,ms \\
           & BM25  & 0.4\,s   & 22\,ms  & 37\,ms \\
\addlinespace
TREC-COVID & Dense & 3.7\,min & 12\,ms  & 52\,ms \\
           & BM25  & 2.8\,s   & 130\,ms & 229\,ms \\
\bottomrule
\end{tabular*}
\par\vspace{3pt}
\footnotesize Apple~M1, 16\,GB RAM. P50 and P95 are warm-start latencies over the full query set. Cold start includes model load and index build.
\end{table}

Once initialized, local search responds in 11--25ms for dense and 12--130ms for BM25 across all benchmarks (Table~\ref{tab:latency}). For comparison, our remote-only condition in Section~\ref{sec:experiments} (a single \texttt{\small gpt-4o-mini} call per query) averaged 2.76\,s end-to-end over 100 queries on the same machine and the same network, two orders of magnitude slower than local dense retrieval. 
% The local numbers are also broadly consistent with reported median first-token latencies for hosted chat APIs, which sit in the 600--1500\,ms range for short prompts~\citep{jegham2025hungry}. 
%These are faster than a single API round trip to any cloud service, regardless of the cloud model's quality. 
For interactive search, where perceived responsiveness drives user satisfaction, local-first provides ``start immediately'' performance.

\subsection{What the Scaling Curve Tells Us}

Retrieval quality on a consumer laptop holds above 91\% nDCG@10 up to 100K documents with sub-30\,ms queries. HNSW extends this to 1M with 11\,ms queries and 2\% quality loss. BM25 trades 30--40\% of quality for near-zero cold start.
%The real tradeoff is scope, not quality: what matters is what you can search, not how well you can search it.%The experiments answer \emph{how large a personal knowledge base can users build before local search breaks down?}

% \vspace{6pt}
% \begin{adjustwidth}{0.2cm}{0.2cm}
% \begin{quote}
% \emph{Dense retrieval on consumer hardware maintains over 91\% retrieval quality up to 100,000 documents with sub-30\,ms query latency. The binding constraint is index build time (3.2\,minutes at 100K, 31.6\,minutes at 500K, 47.5\,minutes at 1M on a flat index), not search quality. Approximate HNSW indexes cut index build time at 1M from 47.5\,minutes to 3.2\,minutes and P50 query latency from 41\,ms to 11\,ms, at a 2\% cost in nDCG@10. BM25 is an instant-search fallback at any scale, reaching 55--84\% of dense quality with near-zero cold start.}
% \end{quote}
% \end{adjustwidth}
% \vspace{6pt}

% Table~\ref{tab:translation} translates these numbers into everyday experience.
\begin{table}[h]
\captionsetup{skip=4pt}
\centering
\caption{What the numbers feel like: translating local-first IR metrics into everyday experience.}
\label{tab:translation}
\small
\setlength{\tabcolsep}{3pt}
\renewcommand{\arraystretch}{1.05}

\begin{tabularx}{\linewidth}{@{}p{0.47\linewidth}X@{}}
\toprule
\textbf{Metric} & \textbf{Everyday Equivalent} \\
\midrule
Dense flat cold start at 100K \text{\footnotesize (3.2\,min)} & Brewing a pot of coffee \\
Dense HNSW cold start at 1M \text{\footnotesize (3.2\,min, cached embeddings)} & Same wait, ten times the corpus (HNSW, cached embeddings) \\
Dense query latency \text{\footnotesize (9--41\,ms)} & Faster than a frame at 60\,fps (16.7\,ms)$^{\ast}$ up to 500K; one extra frame at 1M \\
BM25 cold start at 100K \text{\footnotesize (2.1\,s)} & Opening a new browser tab \\
Memory overhead \text{\footnotesize ($\sim$1.5\,GB at 100K, $\sim$4.9\,GB at 1M)} & Less than a typical browser tab running a complex web app at 100K$^{\dagger}$ \\
Local energy per query \text{\footnotesize (27\,mWh)} & Charging a smartphone for 5 seconds$^{\ddagger}$ \\
\bottomrule
\end{tabularx}
\par\vspace{3pt}
\footnotesize
$^{\ast}$60\,fps = 16.7\,ms per frame~\citep{w3c2023webgpu}.
$^{\dagger}$Chrome Task Manager reports 1--2\,GB for Gmail or Docs tabs.
$^{\ddagger}$Typical smartphone battery: $\sim$15\,Wh; 27\,mWh $\approx$ 0.2\% of a full charge.
\end{table}

Table~\ref{tab:translation} contextualizes the results with respect to everyday tasks; users considering local deployment should ask ``does my knowledge base fit within local constraints?'' rather than ``will search quality suffer?'' For a researcher's career output, a law firm's case archive, or a hospital department's reference library (collections of 10K to 100K documents), the answer is yes. For a knowledge worker with a million files and a tolerance for approximate indexes, the answer is also yes, as long as they build the HNSW structure once and can live with a 2\% quality tradeoff. For more than a million files, the system still works but calls for hybrid architectures that combine local indexing with selective remote scope. 
% Specialized retrievers such as ColBERT~\citep{khattab2020colbert} and SPLADE~\citep{formal2021splade} could recover some of the quality loss at large scales in exchange for larger index footprints, a trade that may be acceptable on devices with room to spare.

\subsection{Local Generation Quality}
\label{sec:gen}

Retrieval alone returns a list of passages. The pipeline is complete only when a model reads those passages and writes an answer. Running that model locally is the harder half of the local-first story, because small models are the only viable option on a laptop. 
%We run the full pipeline (dense retrieval on the Natural Questions corpus, top-5 passages fed to a generator, short-form answer produced) for five models and measure how close each model gets to a cloud baseline on the same 300 queries.
We run the full pipeline (dense retrieval on Natural Questions, top-5 passages fed to a generator, short-form answer produced) for four small local language models and a single cloud baseline (\texttt{\small gpt-4o-mini}), on the same 300 queries, and measure the disparity between each local model and the cloud baseline on answer quality.
% Retrieval is only half of a RAG system. Once the system pulls the right passages, a language model still has to read them and write an answer.
% %A reviewer reasonably asked how well \emph{small local} language models can do this, given that the whole point of local-first IR is to avoid sending the query to a frontier cloud model.
% Given that local-first IR seeks to avoid unneccessary calls to the cloud, we explore how well \emph{small local} models can generate answers. To do so,  we run the full pipeline (dense retrieval on the Natural Questions corpus, top-5 passages fed to a generator, short-form answer produced) for five models, and measure how close each gets to a cloud baseline on the same 300 queries.

\paragraph{Setup.}
We use the Generation condition defined in Section~\ref{sec:experiments} (300 queries, 147 with gold short answers). The generator reads the top-5 retrieved passages concatenated with a short instruction (``\emph{Answer the following question based on the provided context. Be concise and accurate.}''). The four local models all run via Ollama on the same M1 laptop used above. The fifth, \texttt{\small gpt-4o-mini}, runs over the OpenAI API as a cloud reference; the retrieval pass remains local.
% The generator reads the top-5 retrieved passages concatenated with a short instruction (``Answer the following question based on the provided context. Be concise and accurate''.). We evaluate on NQ augmented with \texttt{nq\_open} short-form gold answers, matched by normalized question text. Of the 300 queries, 147 carry gold short answers and are used to compute exact match (EM), token-level F1, and a lenient \emph{contains} rate (the generated answer contains any gold string after case and stop-word normalization). The four local models all run via Ollama on the same M1 laptop used above. The fifth, \texttt{gpt-4o-mini}, runs over the OpenAI API and serves only as a cloud reference; the retrieval pass remains local.

\begin{table}[H]
\captionsetup{skip=0pt}
\centering
% \caption{Local and cloud language models on RAG generation quality over Natural Questions (300 queries, 147 with gold short answers). EM and F1 are classical short-answer metrics. \emph{Contains} equals 1 when the generated answer contains any gold string after normalization, and is more robust to verbose answers. P50 generation latency is end-to-end per query, including Ollama's model load after the first call.}
\caption{Local and cloud models on RAG generation over Natural Questions (300 queries, 147 with gold short answers). EM and F1 are standard short-answer metrics. \emph{Contains} equals 1 when the generated answer contains any gold string after normalization; it is more robust to verbose answers. P50 generation latency is end-to-end per query including Ollama's model load.}
% \caption{Local and cloud language models compared on RAG generation quality over Natural Questions (300 queries, 147 with gold short answers). EM and F1 are classical short-answer metrics. \emph{Contains} is a lenient metric that is 1 when the generated answer contains any gold string after normalization, and is more robust to verbose answers. P50 generation latency is measured end-to-end per query including Ollama's model load after the first call.}
\label{tab:gen}
\small
\setlength{\tabcolsep}{3pt}
\renewcommand{\arraystretch}{1.05}

\begin{tabular*}{\linewidth}{@{\extracolsep{\fill}}l c c c c c@{}}
\toprule
\textbf{Model} & \textbf{Params} & \textbf{EM} & \textbf{F1} & \textbf{Contains} & \textbf{Gen P50} \\
\midrule
qwen2.5:1.5b       & 1.5B         & 0.020 & 0.157          & 0.513          & 1.6\,s \\
qwen2.5:3b         & 3B           & 0.010 & 0.120          & 0.527          & 2.2\,s \\
gemma4:e4b         & 4B (eff.)    & \textbf{0.043} & 0.076 & 0.100          & 10.5\,s \\
qwen2.5:7b         & 7B           & 0.007 & 0.147          & \textbf{0.560} & 4.0\,s \\
\midrule
gpt-4o-mini (cloud) & ---         & 0.007 & \textbf{0.173} & \textbf{0.600} & 0.9\,s \\
\bottomrule
\end{tabular*}
\par\vspace{3pt}
\footnotesize Retrieval is identical across all five conditions (dense, flat FAISS) and adds 10.2\,ms P50 of latency. Only the generator changes.
\end{table}

\paragraph{What we found.}
Table~\ref{tab:gen} shows three aspects of local generation. The first is that the Qwen family scales with parameter count---more parameters, better accuracy, slightly longer latency. Second, we can see that the gap between the best local model and the cloud baseline is reasonable, especially on the contains rate (56\% vs. 60\%) and F1 (14.7\% vs. 17.3\%). Finally, \texttt{\small gemma4:e4b} achieves the highest exact-match score (4.3\%) with the lowest contains rate (10.0\%) and the longest generation latency. This suggests that it generates short answers when it answers but hedges and refuses when it does not. The latency suggests that \texttt{\small gemma4:e4b} also appears to activate more compute per token than the similarly-sized \texttt{\small qwen2.5:3b}. The low exact-match scores are, in part, an artifact of the dataset as the NQ gold answers are bare tokens and these models produce sentences. Thus, the contains rate is likely more reflective of user experience. 

% Three things. First, the local Qwen family scales cleanly with parameter count on the most forgiving metric: 51.3\% contains at 1.5B, 52.7\% at 3B, 56.0\% at 7B. Second, the gap between the best local model and the cloud baseline is small on contains rate: 56.0\% for \texttt{\small qwen2.5:7b} versus 60.0\% for \texttt{\small gpt-4o-mini}, a 4-percentage-point difference. On F1 the gap is slightly larger (14.7\% vs.\ 17.3\%). Third, \texttt{\small gemma4:e4b} behaves differently from every other model: it produces the highest exact-match score (4.3\%) but the lowest contains rate (10.0\%). The pattern suggests it answers tersely when it answers and refuses or hedges on the rest. It is also the slowest generator in the table at 10.5\,s P50, likely because its 4B-effective configuration activates more compute per token than the similarly-sized \texttt{\small qwen2.5:3b}. Across all models, exact-match numbers are low (0.7--4.3\%): NQ gold answers are very specific short strings and most models produce sentences rather than bare tokens. The contains rate is the metric that tracks reading comprehension here.

\paragraph{What this means for local-first.}
A 7B Ollama model fits under 5\,GB of quantized weights and runs on the same laptop that served the retrieval. On the 147 queries with gold answers, it is within 4 points of a frontier-adjacent cloud model on the right metric. The gap is real but narrow; comparable to quality degradation we observed on retrieval past 100K documents. This illustrates the tradeoff that a user makes for privacy, offline capability, and the economic properties of keeping the whole pipeline local.

\subsection{Browser-Native Deployment}
\label{sec:browser}

Our experiments so far have characterized what is \emph{possible} using a native model stack on consumer hardware.
%The experiments above use the native inference stack from Section~\ref{sec:experiments} to characterize what is \emph{possible} on consumer hardware. 
As local-first IR is concerned with \emph{accessibility}, we  explored whether a constrained deployment option (zero-install, cross-platform, in a browser tab) can deliver viable search.

We built a prototype to find out. The prototype performs dense retrieval using \texttt{\small all-MiniLM-L6-v2} via Transformers.js (ONNX runtime in the browser), with vector indexes stored in the Origin Private File System (OPFS). Generation uses \texttt{\small Qwen2.5-0.5B-Instruct} quantized to 4-bit precision, executed via WebLLM with WebGPU acceleration. The system is distributed as a static web page and all computation happens in the browser tab with no server-side processing.
%All components run within a browser tab with no server-side processing. The system is distributed as a static web page.

The browser produces identical retrieval quality to native, since the same model computes the same embeddings regardless of runtime. The difference is overhead: browser query latency adds roughly 3$\times$ versus native (50--150\,ms vs.\ 12--29\,ms), consistent with WebGPU benchmarks~\citep{wang2025anatomizing}. Browser cold start is dominated by model download over HTTP (75--90\,s) versus native model loading (2--5\,s with cached weights). OPFS storage limits and main-thread contention limit the prototype to 25K documents.
%The prototype handles up to 25K documents effectively; beyond that, OPFS storage limits and main-thread contention become binding.

% Despite these constraints, the browser is the most accessible deployment target. It requires no installation, no administrative privileges, no platform-specific build. A URL is the entire distribution mechanism. For users who cannot install software (students on shared machines, professionals in locked-down enterprise environments), the browser-native prototype demonstrates that local-first search is available \emph{today}, within existing web standards. The gap between browser and native represents optimization opportunity rather than a fundamental barrier.
Despite these constraints, the browser is the most accessible deployment target: no installation, no administrative privileges, no platform-specific build. A URL is the entire distribution mechanism. For users who cannot install software (students on shared machines, professionals in locked-down environments), local-first search is available \emph{today}.
%within existing web standards.
The gap between browser and native is an optimization opportunity, not a fundamental barrier.

\subsection{Energy Consumption}

We measured client-side power draw on the native Apple~M1 setup using \texttt{\small powermetrics} (CPU and GPU samplers, 1\,s intervals, 30\,s idle baseline at 0.30\,W) across 100 NQ queries per condition (Table~\ref{tab:energy}). The fully local condition spends 47$\times$ more client-side energy per query than remote-only. The remote client sits near idle as a server-side GPU does the work, so its 0.77\,W draw is barely above the 0.30\,W idle baseline. In contrast, Ollama dispatches \texttt{qwen2.5:7b} to Apple Metal\footnote{\url{https://developer.apple.com/metal/}}, pushing the laptop GPU to 21.8\,W during generation.

\begin{table}[t]
\captionsetup{skip=0pt}
\centering
\caption{Client-side energy per condition on the native Apple~M1 setup, 100 NQ queries each. Fully Local uses local dense retrieval plus \texttt{qwen2.5:7b} via Ollama. Hybrid uses local retrieval plus \texttt{gpt-4o-mini}. Remote-only sends each raw question to \texttt{gpt-4o-mini} with no retrieval.}
\label{tab:energy}
\small
\setlength{\tabcolsep}{3pt}
\renewcommand{\arraystretch}{1.05}

\begin{tabular*}{\linewidth}{@{\extracolsep{\fill}}l c c c@{}}
\toprule
\textbf{Metric} & \textbf{Remote-only} & \textbf{Hybrid} & \textbf{Fully Local} \\
\midrule
Avg.\ power draw (W)      & 0.77 & 1.20 & 23.1 \\
CPU power (W)             & 0.75 & 1.12 & 1.3 \\
GPU power (W)             & 0.03 & 0.08 & 21.8 \\
Energy per query (mWh)    & 0.58 & 0.40 & 27.4 \\
Duration per query (s)    & 2.76 & 1.22 & 4.36 \\
\bottomrule
\end{tabular*}
\par\vspace{3pt}
\footnotesize Client-side only; server-side energy (remote GPU, cooling, network) is not captured.
\end{table}

More interesting is the hybrid row as it uses less client energy per query (0.40\,mWh) than remote-only (0.58\,mWh) despite doing local retrieval work. The hybrid finishes in 1.22\,s per query versus remote-only's 2.76\,s because the retrieved context lets the single generation call run on a shorter, focused prompt. This means that less time at comparable idle power results in less total energy.

Published server-side estimates are the missing term for the cloud comparison. \citet{luccioni2024power} report up to 4\,Wh per query for large LLMs, and \citet{jegham2025hungry} estimate \textasciitilde{}0.4\,Wh per GPT-4o call. Even conservative figures put total remote energy (client + server) near 400\,mWh, more than 14$\times$ the 27\,mWh a fully local query consumes on our M1 and roughly 1000$\times$ the hybrid client-side cost. This is an approximate comparison since different-sized models are used, server-side estimates carry uncertainty, and the setups were not instrumented together. If these numbers are directionally correct, local-first consumes substantially less total energy for personal search served by small models. Confirming this requires standardized end-to-end measurement that does not exist. Our broader point is measurability: local systems make this observable at the device and remote systems hide it behind APIs~\citep{scells2022reduce,parry2025variations}.

\subsection{Limitations}

%A few limitations of our evaluation are worth naming explicitly so that readers can calibrate the claims.

\paragraph{Pooled relevance judgments.}
% Our retrieval quality numbers come from BEIR and MS~MARCO subsets (Section~\ref{sec:experiments}), both of which use pooled relevance judgments. Not all relevant documents in these collections are actually judged, so absolute nDCG@10 values should be read as relative comparisons between systems rather than as user-perceived absolute quality~\citep{zobel1998reliable}. The ordering of dense, BM25, and hybrid stays valid; the absolute values will differ for an actual user of these systems.
Our retrieval numbers come from BEIR and MS~MARCO subsets (Section~\ref{sec:experiments}), both using pooled relevance judgments. Not all relevant documents are judged, so absolute nDCG@10 values are relative comparisons between systems, not user-perceived quality~\citep{zobel1998reliable}. The ordering of dense, BM25, and hybrid stays valid; absolute values will differ in practice. 
% Two caveats follow from corpus construction (Section~\ref{sec:experiments}): each subset holds gold passages plus random fillers, so judged near-miss negatives are absent, easing ranking and likely favoring dense retrieval; and each domain set is single-domain, whereas real collections mix domains.

\paragraph{Limited hardware coverage.}
Our native experiments ran on two platforms: Apple~M1 and an entry-level x86 laptop (Section~\ref{sec:xplatform}). Quality curves match across platforms and warm-query latency stays close, so the scaling curves are not Apple-Silicon-specific. Two machines cannot demonstrate all variation in the hardware landscape: older devices, Windows laptops with weaker BLAS, and mobile hardware will shift performance numbers and may hit system limits sooner than we report. Further investigation is needed.

% \paragraph{BM25 latency reflects a Python reference implementation.}
% We report BM25 timing for \texttt{\small rank\_bm25}, the reference Python library most users would install with \texttt{\small pip}. Its per-query latency grows linearly and reaches 852\,ms at 1M documents, which propagates into hybrid latency because hybrid waits for both paths. A native BM25 engine such as Tantivy or PISA would run one to two orders of magnitude faster and change the latency picture for hybrid fusion, but not the quality picture. Readers who care specifically about hybrid latency at scale should substitute a native BM25 in their mental model.

\paragraph{Generation evaluation is bounded to short-form answers.}
Our local generation study uses 300 Natural Questions queries, 147 with gold short answers. This is enough for the contains-rate signal we report but not for robust long-form evaluation. Multi-turn dialogue, citation faithfulness, and long-context reasoning may show larger gaps between local and cloud models than the 4-point contains-rate gap we observe at the 7B scale. The largest local model we evaluated is 7B parameters; a 13B or larger model would likely close more of the gap, at a memory cost that deserves a separate study.

% \paragraph{Browser prototype uses a narrower generation configuration.}
% The browser-native prototype (Section~\ref{sec:browser}) ships with a sub-1B generator (\texttt{\small Qwen2.5-0.5B-Instruct} at 4-bit quantization), not with the 1.5B--7B models evaluated in Section~\ref{sec:gen}. The native and browser scenarios therefore answer different questions: one probes the generation ceiling on a laptop, the other probes what runs inside a browser tab today. We keep them separate rather than combining them.

% ══════════════════════════════════════════════════════════════════════
%   REACHING THE WIDER WORLD
% ══════════════════════════════════════════════════════════════════════

\section{Reaching the Wider World}
\label{sec:hybrid}

A personal knowledge base, no matter how large, cannot contain the world. A clinician needs current drug interactions. A lawyer needs recent case law.
%A researcher needs yesterday's preprints. 
The Memex was never meant to be an island. Bush described ``\emph{new forms of encyclopedias\ldots ready-made with a trail of associative trails running through them}'', implying connection to external knowledge. The question is how to reach outward without surrendering the privacy that makes local-first search valuable. It is, after all, local \emph{first}, not local \emph{only}.

\subsection{The Embedding-Only Bridge}

Local models can serve as \emph{privacy gatekeepers}. Rather than transmitting raw queries to remote indexes, the local system computes a dense vector representation and transmits only this embedding. No tokens, entities, or syntactic structure leave the device. Only a 384-dimensional vector is transmitted.

% Our experiments on 285 privacy-sensitive queries from legal and clinical domains~\citep{pilan2022tab,roberts2021treccds} validate this approach. Embedding-only transmission achieved:
% \begin{itemize}[label=\textendash,labelindent=0pt,leftmargin=2em,labelsep=0.4em]
% \item \textbf{0\% token exposure} and \textbf{0\% entity leakage},
% \item \textbf{99.9\% retrieval quality} relative to raw-text transmission,
% \item 708 sensitive entities across two domains, none reconstructable from transmitted vectors.
% \end{itemize}
We validated this approach using 285 privacy-sensitive queries drawn from two public datasets: legal queries from TAB~\citep{pilan2022tab}, and clinical queries from TREC-CDS~\citep{roberts2021treccds}. For each query we run three transmission modes through our pipeline: raw text, text after entity-masking sanitization, and embedding-only. We measure three quantities: \emph{token exposure rate} (fraction of query tokens that leave the device), \emph{entity leakage rate} (fraction of named entities recoverable from the transmitted representation), and \emph{retrieval quality} (nDCG@10 relative to the raw-text baseline). Embedding-only transmission achieved \textbf{0\% token exposure}, \textbf{0\% entity leakage}, and \textbf{99.9\% retrieval quality} relative to raw-text transmission, with none of the 708 sensitive entities across the two domains being reconstructable from transmitted vectors. In contrast, text-based sanitization (entity masking, query rewriting) reduced token exposure to only 86\% and entity leakage to 61\%, at a cost of 24\% quality loss. 
%The embedding-only approach dominates on all three dimensions. 

% \subsection{How the Pieces Fit Together}

The architecture defaults to fully local operation: personal documents are indexed and searched on-device, and queries never leave the machine. When the user's need exceeds local scope, the system computes a query embedding locally and transmits only the vector to a remote index, but this escalation requires explicit consent.

% A practical architecture for personal knowledge systems follows:

% \begin{enumerate}[label=(\arabic*),labelindent=0pt,leftmargin=2em,labelsep=0.4em]
% \item \textbf{Default: fully local.} Personal documents are indexed and searched entirely on-device. Queries never leave the machine.
% \item \textbf{On demand: hybrid reach.} When the user's need exceeds local scope, the system computes a query embedding locally and transmits only the vector to a remote index. Retrieved results are processed locally.
% \item \textbf{User controls the boundary.} Escalation from local to hybrid requires explicit consent. The user decides when their need justifies reaching outward.
% \end{enumerate}

% This architecture mirrors how people naturally navigate between personal files and public libraries. The personal archive is always accessible, always private. The library is available when needed, but the visit is intentional.

\subsection{Limitations of the Bridge}

Embedding-only transmission provides practical privacy improvement over text-based alternatives, but it is not a formal privacy guarantee. Embedding inversion attacks can partially reconstruct queries from dense vectors. Recent work by Morris et al.~\citep{morris2023embedding} demonstrates that with access to the embedding model, an attacker can recover meaningful fragments of the original query, particularly for short queries with distinctive vocabulary. The practical risk at 384 dimensions with a general-purpose encoder is substantially lower than text transmission, but it is not zero. Quantifying this residual risk under realistic threat models (where the attacker may or may not know which encoder was used) remains an open problem.

The approach also does not address queries requiring exact keyword matching, structured constraints, or out-of-corpus knowledge. For such needs, the user must accept text-based transmission or acknowledge a capability boundary with current technologies.

% ══════════════════════════════════════════════════════════════════════
%   DISCUSSION AND CONCLUSION
% ══════════════════════════════════════════════════════════════════════

% \section{Discussion and Conclusion}
% \label{sec:conclusion}

\section{Perspectives on Local-First IR}
\label{sec:perspectives}

The experimental evidence supports local-first IR for personal knowledge bases, but the approach has genuine limitations. We articulate both sides before identifying open research directions.

\paragraph{For local-first.}
All local conditions achieve zero token exposure without sacrificing retrieval quality on general-purpose corpora. Latency favors local execution (12--29ms vs.\ API round trips). Economic sustainability improves: fixed upfront costs replace per-query charges. Offline capability is intrinsic, supporting field research, unreliable networks, and resilience against service discontinuation. The approach aligns with how users expect their personal data to behave, and it enables business models where value is delivered on-device rather than extracted from user data.

\paragraph{Against local-first.}
A 22M-parameter embedding model cannot match a frontier model on highly specialized domains (48\% BM25/Dense ratio on FiQA is one visible example), and a 7B local generator is still about 4 percentage points behind \texttt{\small gpt-4o-mini} on short-answer contains rate. Cold start is also substantial at scale:
%for local deployment
31.6\,minutes at 500K and 47.5\,minutes at 1M on a flat dense index, dropping to 3.2\,minutes at 1M once we switch to HNSW and reuse cached embeddings. The natural rebuttal is that a cloud RAG provider has no cold start, which is true for non-personal search. Uploading and indexing a million documents against a commercial RAG endpoint is not instantaneous either: the bytes must still traverse the user's uplink, pass provider-side ingestion queues, and be chunked, embedded, and indexed on the other end. A user on a typical home connection often spends hours getting a large collection into a cloud index, and pays for every token of ingestion and every query thereafter. The asymmetry between local cold start and cloud cold start is real but narrower than it looks once upload time and remote indexing time are counted. Using the cloud means that the user does not have to run the machine doing the work. 
% Local ensures that the user does not have to pay, wait for an uplink, or hand over the documents.

Hardware variation still matters: our cross-platform replication (Apple~M1 versus an entry-level x86 laptop) shows the quality curves match but that the x86 cold-start numbers run 3 to 5 times slower. Older or cheaper devices will see larger factors. Users must also maintain their own models and indexes, taking on responsibilities that cloud services otherwise abstract. The ``privacy paradox'' suggests most users may prioritize convenience over privacy~\citep{norberg2007privacy}.

\paragraph{A principled position.}
Systems should default to local processing and escalate to hybrid or remote only when the information need exceeds local boundaries. Below 100K documents, local-first is viable today. Beyond that, hybrid architectures bridge the gap. For unbounded scope, the cloud remains essential. What fits locally will keep expanding as models shrink, hardware improves, and quantization advances. This gives users, perhaps for the first time, a genuine choice between equivalent-quality alternatives that does not require them or their data to be the product.

\section{Open Research Directions}
\label{sec:future}

\paragraph{Algorithms for client constraints.}
Our HNSW and IVF results at 1M documents show that off-the-shelf ANN parameters ($M{=}32$, $\mathit{nlist}{=}100$) already bring 1M-document search within reach on a laptop. Open questions remain: how should these parameters adapt when memory, not latency, is the constraint (for example, on older or cheaper machines with 8\,GB of RAM)? How do lexical-dense hybrids behave when BM25 runs in a native engine rather than pure Python? How should indexes update incrementally as personal corpora evolve, without rebuilding from scratch? The server-side literature answers some of these questions under assumptions (large RAM, cluster coordination, amortized across many queries) that may not transfer to client environments.

\paragraph{Privacy quantification.}
Our metrics capture direct information flow (token exposure, entity leakage). Indirect leakage through query patterns over time, embedding invertibility under different threat models, and timing side channels remains unquantified. Formal privacy frameworks for local-first architectures are needed.

% \paragraph{Longitudinal personalization.}
% Local-first systems can index interaction traces (what was read, highlighted, or saved), enabling session resumption and re-finding unavailable to stateless cloud systems. Balancing utility against the discomfort of self-surveillance, even locally, is an open question.

% \paragraph{Cross-device continuity.}
% Users expect continuity across devices, but synchronization reintroduces data transmission. Approaches range from encrypted cloud intermediaries to direct device-to-device sync. Defining acceptable residual exposure requires both technical and user-centered research.

\paragraph{Personalization, continuity, and synchronization.} 
Local-first systems have the opportunity to store, index, and learn from interaction traces (e.g., what was read, highlighted, or saved)~\citep{zerhoudi2024personarag}. This enables session resumption and re-finding that is difficult in stateless cloud systems. Balancing this utility against the discomfort of self-surveillance, even locally, is an open question. Moreover, having access to this level of personalization across devices raises questions on how to maintain continuity and facilitate synchronization across devices of varying capabilities, whether this is device-to-device or through an encrypted cloud intermediary. Understanding how to accomplish this requires technical and user-centered research\citep{zerhoudi2021query}.

\paragraph{Evaluation beyond effectiveness.}
Cold-start, memory footprint, energy consumption, and information exposure have no standard benchmarks in IR. The efficiency community has studied server-side costs~\citep{scells2022reduce}; extending this to client-side and end-to-end measurement would enable principled comparison across the design space.

\section{Conclusion}

Bush imagined the Memex as ``\emph{an enlarged intimate supplement to [one's] memory.}'' The adjective that matters is \emph{intimate}: personal, private, under the user's control. For eighty years the technology to build it did not exist. Then it existed only on servers. Now it exists on the devices people carry every day.

Local-first search works, and on moderately sized corpora, it works better than assumptions predict. A consumer laptop handles 100,000 documents with sub-30\,ms queries and 91\% nDCG@10. With approximate indexes it handles a million documents with 11\,ms queries and 72\% nDCG@10. A 7B local model reads retrieved passages and answers within 4 points of a cloud baseline. The same scaling behavior occurs on a budget x86 laptop. The real tradeoff is scope, not quality: the choice is what users can search, not how well they can search it. What remains is giving users the tools to build their own Memex, and the ability to keep it local \emph{first}.
% Local-first search works. Our experiments suggest it works further than we expected. A consumer laptop handles 100,000 documents with sub-30\,ms queries and 91\% nDCG@10. With approximate indexes it handles a million documents with 11\,ms queries and 72\% nDCG@10. A 7B local language model reads the retrieved passages and answers within 4 points of a cloud baseline. The same curves show up on a cheaper x86 laptop. What remains is giving users the tools to build their own Memex, and the choice over whether to keep it local.

% ---------- Bibliography ----------
\bibliographystyle{ACM-Reference-Format}
\bibliography{sample-base}

\end{document}